\documentclass{emulateapj}

\shorttitle{TA-DA}
\shortauthors{Da Rio et al. 2012}

\begin{document}

\title{TA-DA: a Tool for Astrophysical Data Analysis}

\author{Nicola Da Rio}
\affil{European Space Agency, Keplerlaan 1, 2200 AG Noordwijk, The Netherlands}
\email{ndario@rssd.esa.int}

\author{Massimo Robberto}
\affil{Space Telescope Science Institute, 3700 San Martin Dr., Baltimore MD, 21218, USA\ }

%%=============================================================================================
%%=============================================================================================
%%=============================================================================================

\begin{abstract}
We present TA-DA, a new software aimed at greatly simplify and improve the analysis of stellar photometric data in comparison with theoretical models, and allow the derivation of stellar parameters from multi-band photometry. Its flexibility allows one to address a number of such problems: from the interpolation of stellar models, or sets of stellar physical parameters in general, to the computation of synthetic photometry in arbitrary filters or units; from the analysis of observed color-magnitude diagrams, to a Bayesian derivation of stellar parameters (and extinction) based on multi-band data. TA-DA is available as a pre-compiled IDL widget-based application; its graphical user interface makes it considerably user-friendly.
In this paper we describe the software and its functionalities.
\end{abstract}

\keywords{Data Analysis and Techniques, Astrophysical Data, Stars, Star Clusters and Associations}

%%=============================================================================================
%%=============================================================================================
%%=============================================================================================

\section{Introduction}
\label{section:introduction}
A common problem in stellar astrophysics is the relation between the observational quantities (magnitudes, colors, fluxes) and the physical parameters of the stars (effective temperature, luminosity, mass, age). Often, dust extinction challenges this task even further, especially when differential reddening affects the individual sources in a region. Whereas it has always been common practice to analyze photometry data in 2-dimensional space, like a color-magnitude diagram (CMD) or a 2-color diagram (2CD), often this approach is insufficient to fully characterize the individual stars. In fact, there are well-known degeneracies among the stellar parameters for stars of given color: a few examples are the effective temperature ($T_{\rm eff}$)-extinction degeneracy for low- and intermediate-mass stars and optical magnitudes \citep[e.g.,][]{hillenbrand97,bailer-jones2011}, or the age-metallicity degeneracy for RGB stars \citep[e.g.,][]{vandenberg2006,crnojevic2010}.

Multi-band photometry often allows to improve upon these limitations, both in breaking the degeneracies and reducing the overall uncertainty. Today, the availability of large field surveys such as the \emph{Sloan Digital Sky Survey} (SDSS, \citealt{abazajian2009}), the \emph{Two-Micron All-Sky Survey} (2MASS, \citealt{skrutskie2006}) and the \emph{Wide-field Infrared Survey Explorer} (WISE, \citealt{wright2010}), the diffusion of public data archives and the rise of Virtual Observatory (VO) infrastructures, enable easy access of an overwhelming quantity of photometric data for the community.

Even with a great variety of data in one's hand, the characterization of the properties of stars may hide further obstacles, some of which merely practical. Various theoretical models predicting the stellar parameters are known to differ from one another, requiring one to assess, by comparison with the data, the level of trustworthiness of the models. Often, stellar evolutionary models already provide the predicted magnitudes and colors; this conversion, however, may not be well calibrated, or not available in the specific photometric bands that one requires. In this case, common solutions are to transform the available photometry from one to another system -- with the consequent uncertainties -- or to rely on synthetic photometry to recompute the predicted fluxes.

\begin{figure*}
\epsscale{1}
\includegraphics[width=17cm]{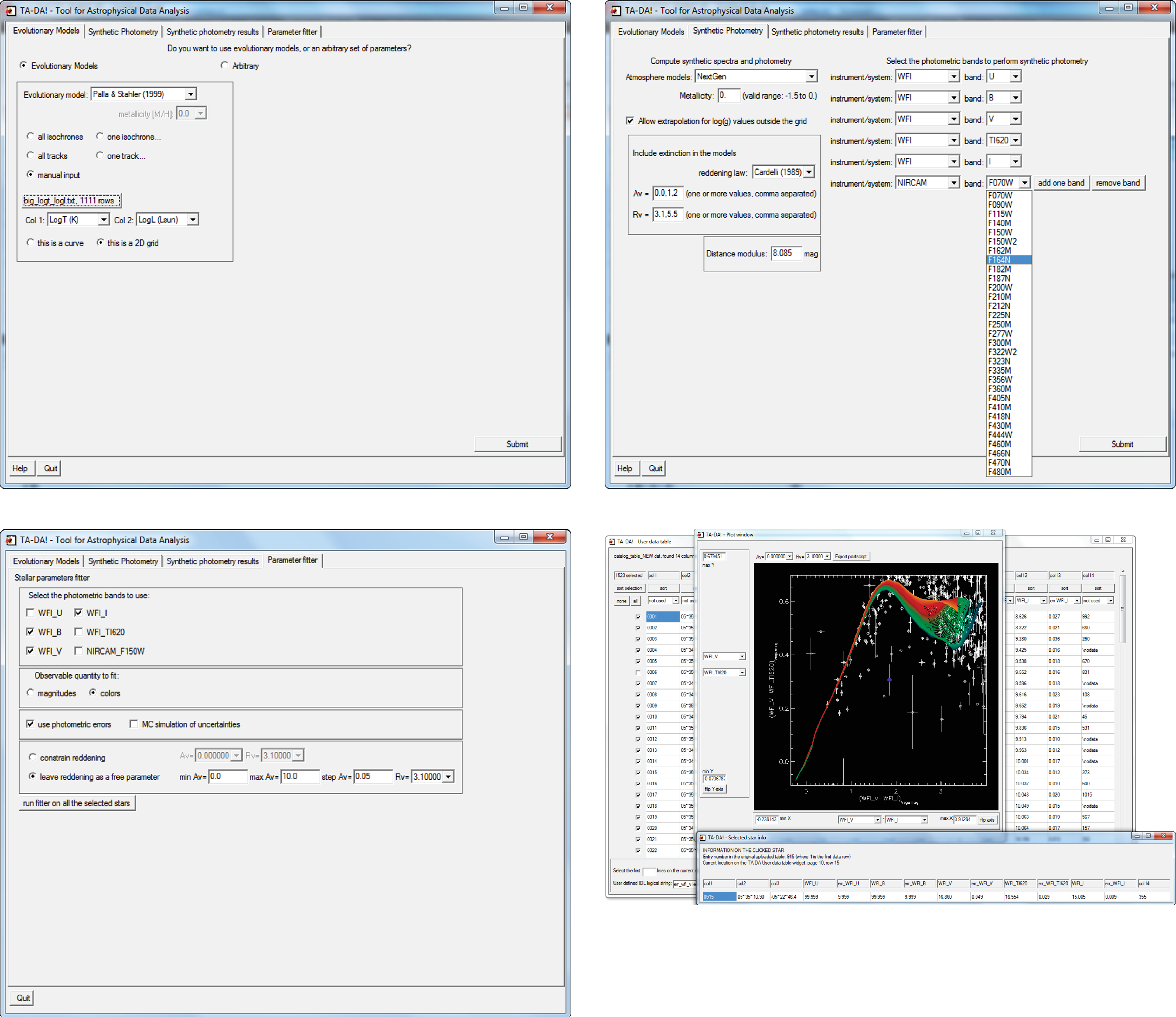}
\caption{Some examples of the graphical user interface of TA-DA: \emph{top left:} the interpolation of evolutionary models or the definition of the model stellar parameters; \emph{top right:} the interface for the synthetic photometry; \emph{bottom right:} the visual inspection of the computed synthetic photometry, the attachment of observed photometry and the selection of sources within the observed catalog; \emph{bottom left:} the interface for the stellar parameter fitter.  \label{fig:screenshots}}
\end{figure*}

There are several available tools to simplify some of these steps. A number of stellar evolutionary models provide online interfaces for interpolation of their grids and computation of magnitudes and colors (e.g, the \citealt{siess2000} pre-main sequence models or the Padova models from \citealt{marigo2008} and \citealt{girardi2010}). Similarly, some grids of synthetic spectra already provide, besides the computed spectra, also tools to compute the integrated colors and magnitudes in several photometric systems (e.g., the \texttt{PHOENIX} models of the Lyon group, see \citealt{allard2011}). For what concern the synthetic photometry in general, a very popular code is \texttt{SYNPHOT}, an IRAF package part of the \emph{Space Telescope Science Data Analysis System} (STSDAS) which performs several different types of such calculations. As for the estimate of stellar parameters and extinctions from multi-band photometry, a powerful code is \texttt{CHORIZOS} \citep{maiz-apellaniz2004}. This software allows one to estimate $T_{\rm eff}$, $\log g$, the extinction $A_V$ as well as the reddening parameter $R_V$ from multi-band colors, using a Bayesian approach. \texttt{CHORIZOS} also performs automatic synthetic photometry, with the possibility of adding custom filter bands. Still, \texttt{CHORIZOS} is designed to fit stellar colors, not magnitudes, neglecting the additional information in the data provided by the luminosities; moreover, stellar evolutionary models are not directly integrated in the code.
All these tools, in fact, have the common shortcoming that either they are designed to address only a particular problem within the general analysis, or that they are constrained to perform as black-boxes.

Following these premises, it appears that a new, integrated, flexible code for photometric data analysis in relation to stellar parameters is needed. With this motivation, we have developed the \emph{Tool for Astrophysical Data Analysis} (TA-DA). In this paper we present the code, its general functionalities, and show some example of use.

\section{Requirements for TA-DA}
\label{section:requirements}

TA-DA is conceived as a tool for quick-look, comparison, and parameter fitting of multi-band photometry in relation to stellar models. In the development of the code, we focused on two general concepts:
\begin{enumerate}
\item \emph{Generality and Versatility}: TA-DA should be able to address a range of different problems concerning the relation between photometry and stellar parameters.
\item \emph{Usability}: TA-DA must be user-friendly, ensuring the users to exploit its functionalities (or test if it is useful for their scientific aims) with little or no extra time dedicated to familiarize with the code.
\end{enumerate}
Specifically, we included the following functionalities:
\begin{itemize}
    \item Integration with stellar evolutionary models: TA-DA is natively able to load and interpolate stellar interior models, allowing one to fit model-dependent quantities such as stellar masses and ages. The consideration of interior models, nevertheless, is not mandatory, as ultimately one could solely consider, e.g., $T_{\rm eff}$ as the only stellar parameter.
    \item Automatic and universal synthetic photometric engine: TA-DA is able to perform synthetic photometry, converting stellar parameters into magnitudes or fluxes in a number of units, allowing also reddening due to dust extinction to be directly applied to the synthetic spectra. The synthetic photometry can be performed on different grids of atmosphere models, and on arbitrary photometric systems.
    \item Data visual quick-look in comparison with models or theoretical stellar parameters: the user can load photometric tables, and plot them together with the pre-computed sets of stellar parameters, constructing, e.g., CMDs or 2-CDs, and varying the amount of assumed dust extinction. This is useful for obtaining a first visual check on the behavior of the data, and refine the selection of sources to be further studied.
    \item Automatic parameter fitter: TA-DA allows one to automatically estimate the stellar parameters of the individual sources from their multi-band photometry; depending on the data at disposal and the number of free parameters, this can be either an exact solution or a probabilistic one. The number of free parameters, as well as the photometric bands to use, can be fully customized.
\end{itemize}

\section{How TA-DA works.}
\label{section:how-tada-works}

TA-DA has been developed using the ITTVIS \emph{Interactive Data Language} (IDL), and presents a graphical user interface (GUI) based on widgets; it is distributed as a pre-compiled, cross-platform code. The distribution package already includes a number of stellar evolutionary and atmosphere models, as well as a full repository of photometric filter throughput profiles. We refer the readers to the Appendix \ref{appendix:models-filters} for additional details.

The TA-DA main GUI is divided into 4 panels, respectively for: a) the specification of the model stellar parameters and the interpolation of evolutionary models; b) the synthetic photometry; c) the visual display of the synthetic photometry, the handling of the observed photometry table, selection and plotting of the latter; d) the stellar parameter fitter. We will now describe some of these functionalities.

\subsection{Stellar parameters and evolutionary models}
\label{section:how-tada-works-evolmodels}

TA-DA includes several families of evolutionary models; for pre-main sequence (PMS) populations we have included models from \citet{baraffe98,dantona98,palla99,siess2000,tognelli2011}, for evolved population we included the Padova models from \citet{marigo2008} with the \citet{girardi2010} correction. If multiple values of metallicity are available for a given set of interior models, a value should be specified. When instead a grid of models includes additional parameters, such as in the case of the PISA models from \citet{tognelli2011}, computed for several values of mixing length parameter $\alpha$, helium enrichment, and deuterium initial abundances, these are treated as different families of models. Additional interior models can be added by a user provided they are stored in the proper format.

TA-DA allows one to consider the entire grid, for a selected metallicity, or to restrict it to a particular arbitrary isochrone or mass track. Furthermore, it is possible to manually input user defined sets of stellar parameters, attaching a table reporting an arbitrary set of 2 quantities chosen from the following 5: mass, $\log$ age, $\log T_{\rm eff}$, $\log L_{\rm bol}$, stellar radius. It should be noted that any combination of 2 of these parameters automatically define the remaining ones, since, e.g, the stellar radius $R$ relates to $T_{\rm eff}$ and $\log L$ through $L=4\pi R^2 T_{\rm eff}^4$ and the evolutionary models natively map the Hertzsprung-Russel Diagram (HRD) into stellar masses and ages. Thus TA-DA will consider the user-supplied parameters and derive the remaining ones, including the surface gravity $\log g$, needed for the synthetic photometry (see below).

Alternatively, TA-DA also allows one to consider only parameters in the HRD ($T_{\rm eff}$, $R$, $\log g$ for a specified metallicity, if available), the minimum ones necessary to both derive the synthetic photometry and allow the fitting of stellar parameters, without relying on model-dependent physical quantities such as stellar masses and ages.

\subsection{The synthetic photometry}
\label{section:how-tada-works-synphot}

TA-DA is able to perform automatic synthetic photometry, i.e., to convert stellar parameters into observed fluxes or magnitudes in a given photometric filter, relying on grids of synthetic spectra. To this purpose, one must select a grid of atmosphere models and, if available, the metallicity [$M/H$]. Dust extinction can be also added (by declaring one or several values of $A_V$), and directly applied to the synthetic spectra \emph{before} the computation of the synthetic photometry. We included the Galactic reddening laws from \citet{cardelli89}, as well as those from \citet{gordon2003} for the LMC average, LMC2 supershell and SMC bar from \citet{gordon2003}. In first case, besides $A_V$ one (or more) values of the reddening parameter $R_V$ must be specified. The distance modulus in magnitudes must also be provided. Finally, one specifies the photometric system and bands in which the synthetic photometry must be performed (see the Appendix \ref{appendix:models-filters} for details).

Synthetic photometry is performed in the standard way, by integrating the synthetic spectra within the filter bandwidths and normalizing onto a spectrum of Vega; the observed magnitude in a photometric band $S_{\lambda}$ of a star with a spectral energy distribution $F_{\lambda}$, stellar radius $R$, and true distance modulus $\mu=(m-M)_0$ is given by
\begin{equation} \label{equation:first}
M_{S_{\lambda}}=-2.5\log\bigg[\bigg(\frac{R}{10\textrm{pc}}\bigg)^{2}\frac{\displaystyle \int_{
\lambda } \lambda F_{\lambda} S_{ \lambda } 10^{-0.4A_{\lambda}}
\textrm{d} \lambda }{\displaystyle \int_{ \lambda } \lambda f^{0}_{\lambda} S_{
\lambda } \textrm{d} \lambda }\bigg]+ZP_{S_{\lambda}}+\mu
\end{equation}
where $f^{0}_{\lambda}$ is a reference spectrum that gives a known apparent magnitude $ZP_{S_{\lambda}}$; in the \textsc{Vegamag} standard, which uses the flux of $\alpha$~Lyr as reference, $f^{0}_{\lambda}=F_{\lambda , {\rm VEGA}}$ and the zero-points $ZP_{S_{\lambda}}$ should be zero by definition, although this is not always the case for some old photometric systems. Finally, $A_{\lambda}$ is the reddening law, scaled to the selected values of $A_V$. Equation~(\ref{equation:first}) can be rewritten then as:
\begin{eqnarray} \label{equation:second} M_{S_{\lambda}} = &
-5\log\bigg( \frac{\displaystyle R_{\odot} }{\displaystyle 10 {\rm pc}
}\bigg) - 5\log\bigg(\frac{\displaystyle R_{\star}}{\displaystyle
R_{\odot}}\bigg) + \mu + B(F_{\lambda},S_{\lambda}) \nonumber \\ = & 43.2337 -
5\log\bigg(\frac{\displaystyle R_{\star}}{\displaystyle R_{\odot}}\bigg)
+ \mu + B(F_{\lambda},S_{\lambda})
\end{eqnarray}
where
\begin{equation}
\label{equation:third}
B(F_{\lambda},S_{\lambda})=-2.5\log\bigg(\frac{\displaystyle \int_{
\lambda } \lambda F_{\lambda} S_{ \lambda } 10^{-0.4A_{\lambda}} \textrm{d} \lambda
}{\displaystyle \int_{ \lambda } \lambda F_{\lambda ,{\rm Vega}} S_{ \lambda
} \textrm{d} \lambda}\bigg)
\end{equation}
The latter term can be directly calculated having the synthetic spectrum for the $T_{\rm eff}$ and $\log g$ corresponding to a given point of the input model parameters, a calibrated Vega spectrum and the band profile. The Vega normalization is performed using a recent reference spectrum of Vega \citep{bohlin2007}. We have considered the zeropoints $ZP_{S_{\lambda}}$
from \citet{maiz-apellaniz2007}, collected from several sources for the code \texttt{CHORIZOS}.

Although the default units for the computed synthetic photometry are Vegamag magnitudes, TA-DA allows to switch also to ABmag, STmag, Jansky or erg s$^{-1}$cm$^{-2}$$\AA^{-1}$. This is performed by computing, for every filter, the flux corresponding to $m=0$~mag by direct integration onto the spectrum of Vega. Then, since ABmag and STmag are related to constant flux densities (respectively per unit of spectral frequency and wavelength), these ``zero-point'' fluxes automatically provide the offsets between the 3 magnitude standards.

TA-DA allows one to save the results of synthetic photometry to file, as well as to use them for the subsequent analysis (visual inspection and parameter fitting). One can also save a table with most of the relevant characteristics of the selected filters. These include, besides the aforementioned zeropoints, also effective wavelength, full width half maximum (FWHM) and equivalent width (EW). This information can be particularly useful when one adds new (non-conventional and not pre-characterized) filters to the tool, as TA-DA automatically derives these quantities on the fly.

TA-DA plots the results of synthetic photometry on screen on a dedicated window. Here one chooses the quantities to be plot, either a magnitude-magnitude plane, a CMD, or a 2CD. The units and extinction to be considered can also be edited in real time. An interesting feature that is helpful for a fast quick look at the result of synthetic photometry is that one can see the parameters associated to individual points of the model grid by clicking directly on the plotting window. All the physical parameters associated to that point, as well as the displayed observational quantity  (e.g., color and magnitude) are then reported in a separate popup window.

\subsection{Observed photometry vs. models}
\label{section:how-tada-works-quicklook}
TA-DA enables one to load a table reporting observed photometry, both for quick look in comparison with the models and for the fit of the stellar parameters. Currently the software supports both ASCII formats and Virtual Observatory table. A new GUI window is created, reporting attached table, and enabling one to browse through the data, sort by columns, and select individual sources. In this panel, one must specify what columns correspond to what filter, as well as the associated uncertainties (if available); the options are chosen among the photometric bands previously selected for the synthetic photometry. The selection of individual sources can be performed both manually (clicking on table entries), or automatically, through logical conditions not restricted to the photometric data. An example could be: select all the source with $V<20$, with the photometric errors $\sigma I<0.1$ and where the last column of the table does not report the string 'Class~I'. The selected sources are also automatically shown in the plotting window of TA-DA, together with the models previously computed. Similarly as for the models, one can click directly on individual stars on the plot, and see on screen the location of that source in the catalog, as well as the value for that star of all the columns in the table.

\subsection{The parameter fitter}
\label{section:how-tada-works-fitter}
With all the necessary preparation completed (a catalog of observed magnitudes or fluxes and an a grid of models converted into the same observables through synthetic photometry), TA-DA is able to perform a series of fitting techniques to estimate the parameters of the observed sources. Specifically, it allows one to choose which bands to consider, and if magnitudes or colors are to be used. The latter options is useful when one aims to derive the stellar parameters in a luminosity (or radius) independent fashion, in a 2-color or multi-color space. Clearly, choosing colors instead of luminosities decreases the dimension of the observational space by one unit.

As for the theoretical space -- the result of synthetic photometry -- in general one could leave from 1 to 3 free parameters to be evaluated by the fitter. One or two parameters correspond to the photosphere. This depends on the model previously assumed for synthetic photometry: for instance, if one previously considered 1 single isochrone, or mass track, or arbitrary curve in the ($T_{\rm eff}$-$\log g$) plane, etc., this will lead to a 1 free dimension in the parameter space. Conversely, considering an entire grid of interior models (masses and ages), or a 2-dimensional grid of $T_{\rm eff}$ and stellar radii brings the free photospheric parameters to be determined to two. The additional free parameter is the extinction $A_V$ for individual stars. TA-DA allows to leave $A_V$ unconstrained within a pre-defined range, as well as to constrain it to a given value for all stars. If synthetic photometry was previously carried out for multiple values of the reddening parameter $R_V$, which is possible if the \citet{cardelli89} reddening law is assumed, the value of $R_V$ to be used by the fitter must be specified. We did not allow to leave both $A_V$ and $R_V$ free, as for example in the tool \texttt{CHORIZOS} by \citet{maiz-apellaniz2004}. This is because, in general, most photometric datasets are not accurate enough to allow a precise disentanglement of the two; moreover inaccuracies in the synthetic spectra would likely induce large deviations in the derived values of $R_V$. %Therefore we believe there are very few feasible applications for an assessment of $R_V$, together with the rest of the parameters, based on photometry.

In order to make the fit possible, the dimension of the observational space $m$ (number of colors or magnitudes)  must be greater or equal than the number of free parameters $n$. Although from the algorithmic point of view there is no separation between the cases $m>n$ and $n=m$, there are some conceptual differences. The first case is analogous to a standard SED fitting, in which the model space is explored (and if needed interpolated to a finer sampling) until the minimum $\chi^2$ is determined (see Equation \ref{equation:chisquare} below); if $\chi^2_{min}\simeq1$, the best solution is also a good fit, otherwise, the data may be incompatible with the models, or the associated uncertainties underestimated. The second case, on the other hand, corresponds to an interpolation problem, for which the solution is expected to have $\chi^2=0$; if $\chi^2>0$, the location of the data point in the observational spaces lies outside the area (or volume) covered by the models.

In detail, the fitter of TA-DA works as follows: let us consider \textit{\textbf{P}} the $n$-dimensional space of the free parameters ($p_1$,...,$p_n$); the synthetic photometry associates to each point in \textit{\textbf{P}} a point $(s_1,...,s_m)$ in the $m$-dimensional space of the observational quantities, \textit{\textbf{S}}. Then, for every observed star, the observed fluxes, colors or magnitudes $(r_1,...,r_m)$ with the associated uncertainties $(\sigma_{r_1},...,\sigma_{r_m})$ are considered and the $\chi^2$ is simply:
\begin{equation}
\label{equation:chisquare}
\chi^2(p_1,...,p_n)=\sum_{i=0}^m\frac{(r_i-s_i)^2}{\sigma_{r_i}^2}.
\end{equation}
\noindent First TA-DA identifies the set of ($p_1$,...,$p_n$) which minimize $\chi^2$, then \textit{\textbf{P}} is locally oversampled to a denser grid \textit{\textbf{P$_2$}}, and  \textit{\textbf{S}} is interpolated accordingly to $\textit{\textbf{S$_2$}}$. The $\chi^2$ is thus recomputed, and a new set of parameters estimated. The process is iterated until a convergence is found, either $\sim0$ for $n=m$ or to a positive value otherwise. The local oversampling of \textit{\textbf{P}} is made easier by the fact that the grid of parameters is always rectilinear in ($p_1$,...,$p_n$). This could be generally, e.g., a rectilinear grid of masses, ages and $A_V$ or any other choice for the original model stellar parameters, for instance a rectilinear grid of $T_{\rm eff}$, stellar radii $R$, and $A_V$. Since, as mentioned earlier, the stellar parameters mass, age, $T_{\rm eff}$, $\log L$, and $R$ are interconnected, for each solution all these quantities will be provided, irrespective of the type of initial parameters provided to TA-DA.

In the case $n=m$, as already mentioned, a $\chi^2>0$ indicates that the data point lies outside the volume covered by the models. In practice, the fact that fitter converges until the $\chi^2$ does not improve by local oversampling within \textit{\textbf{P}} always leads to small positive values of $\chi^2$. A source located outside the model grid, but by less than one unit of photometric error from the edge, will also lead to a $\chi^2<1$. To precisely distinguish if the point is inside or outside the model grid, therefore, TA-DA also determines this condition geometrically, regardless the minimum $\chi^2>0$.
In any case, even for data points incompatible with the models, the closest solution is found, and the associated $\chi^2$ tabulated in the result. The user must therefore check manually the results and decide if there are any solutions to be rejected, based on the specific scientific goals.

The TA-DA fitter also allows one to estimate the uncertainty associated to each parameter of the best fit-solution for the individual stars. This is performed by TA-DA with a Monte Carlo simulation, in which the photometry is displaced according to the photometric errors. This approach may be computationally time consuming, especially in the case of large grids of models and 3 free parameters, or large observational data sets. This method, however, remains the safest for an accurate estimate of the errors; this is because the relation between \textit{\textbf{P}} and \textit{\textbf{S}} provided by the synthetic photometry is often highly non-linear, and symmetric, well-behaved photometric errors in the observational space relate to highly correlated and skewed probability distributions in the model space.

After the completion of the fitting procedure, TA-DA allows one to save the results to file. These include all the best-fit parameters, with the associated uncertainties if available, the minimum $\chi^2$, and a flag parameter indicating if the solution is a good fit or if the observed photometry was incompatible with the models.

\section{Important caveats}
\label{section:caveats}
\subsection{Accuracy of the synthetic spectra}
\label{section:caveats-atmospheres}
It should be beard in mind that the correctness of the fitted parameters is strongly related to the accuracy of the assumed models, e.g. both the evolutionary models and the synthetic spectra. It is well known that different families of evolutionary models differ in their predicted masses and ages from each other, and this is particularly severe for PMS models \citep{mayne2007,hillenbrand2009}; there is also evidence of systematic offsets between the modeled masses of low-mass stars and the measured once \citep[e.g.,][]{hillenbrand2004}. On the other hand, current atmosphere models do not seem to fully reproduce the observed fluxes, especially for low-mass M-type stars and at young ages, both at optical wavelengths \citep[e.g.,][]{dario2009b,dario2010,bell2012} and, although less prominently, in the near infrared \citep{scandariato2012}. Because of these issues, a dataset compared with models in different colors can lead to systematic differences in the derived parameters \citep[e.g.,][]{naylor2002}.

We auspicate that developments in the stellar atmosphere modeling will solve these problems; alternatively, it would be useful to produce empirically calibrated grids of spectra, constructed to match the observed colors. This has been attempted in the past for main-sequence dwarfs and giants (see, e.g., the BaSeL grid from \citealt{lejeune1997}), but such a calibration is still missing for, e.g., PMS stars.

In any case we warn the users of TA-DA to be aware of the accuracy of the synthetic spectra grids to be used in the parameter range relevant for their astrophysical analysis. We also stress that the plotting and quick look capability of TA-DA (e.g., looking at the same isochrones compared to the observed data in different color-color and color-magnitude planes) can be very helpful to pinpoint possible problems with the fluxes predicted by the models, as well as possible errors in the photometric data calibration.

\begin{figure*}
\epsscale{1.1}
\plottwo{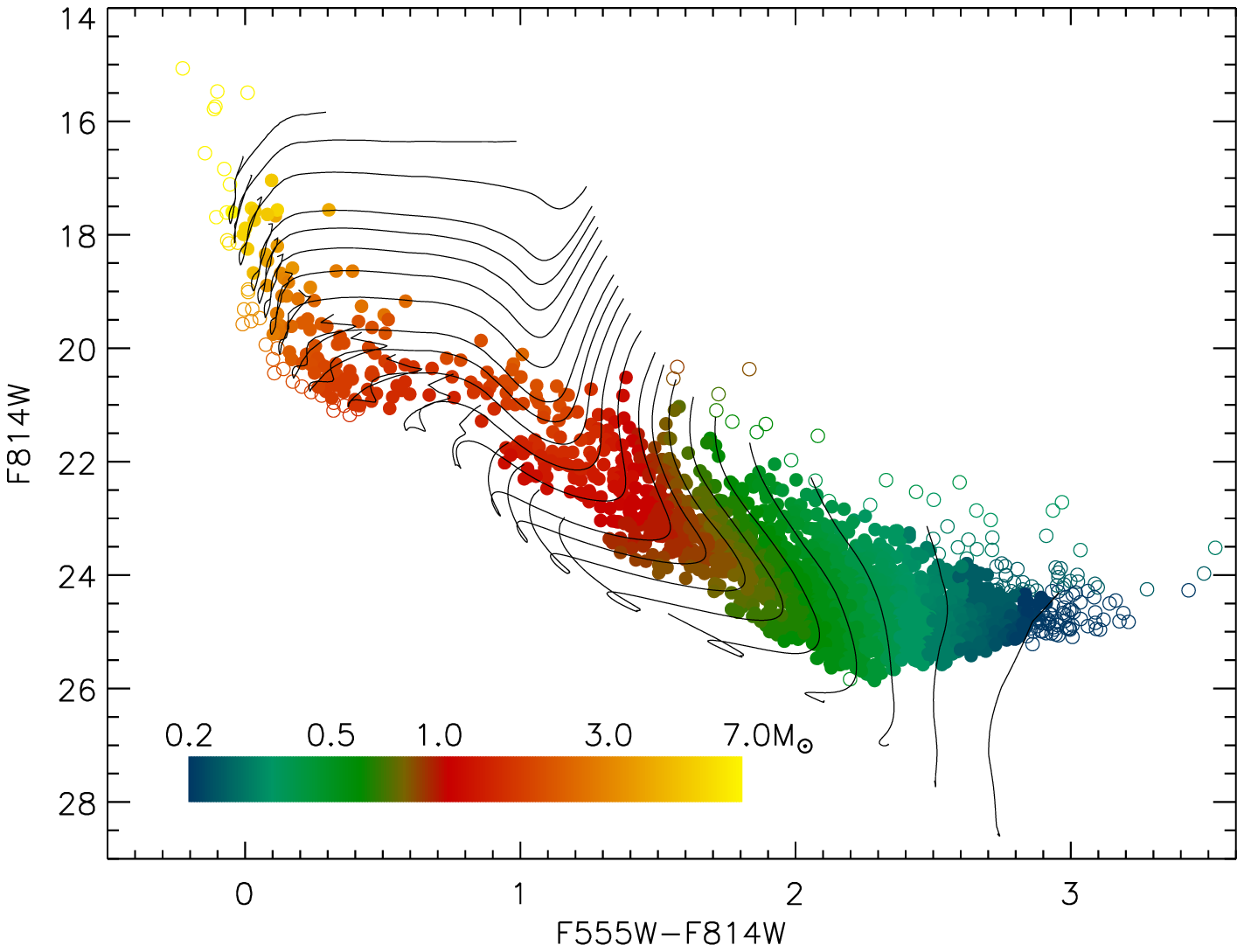}{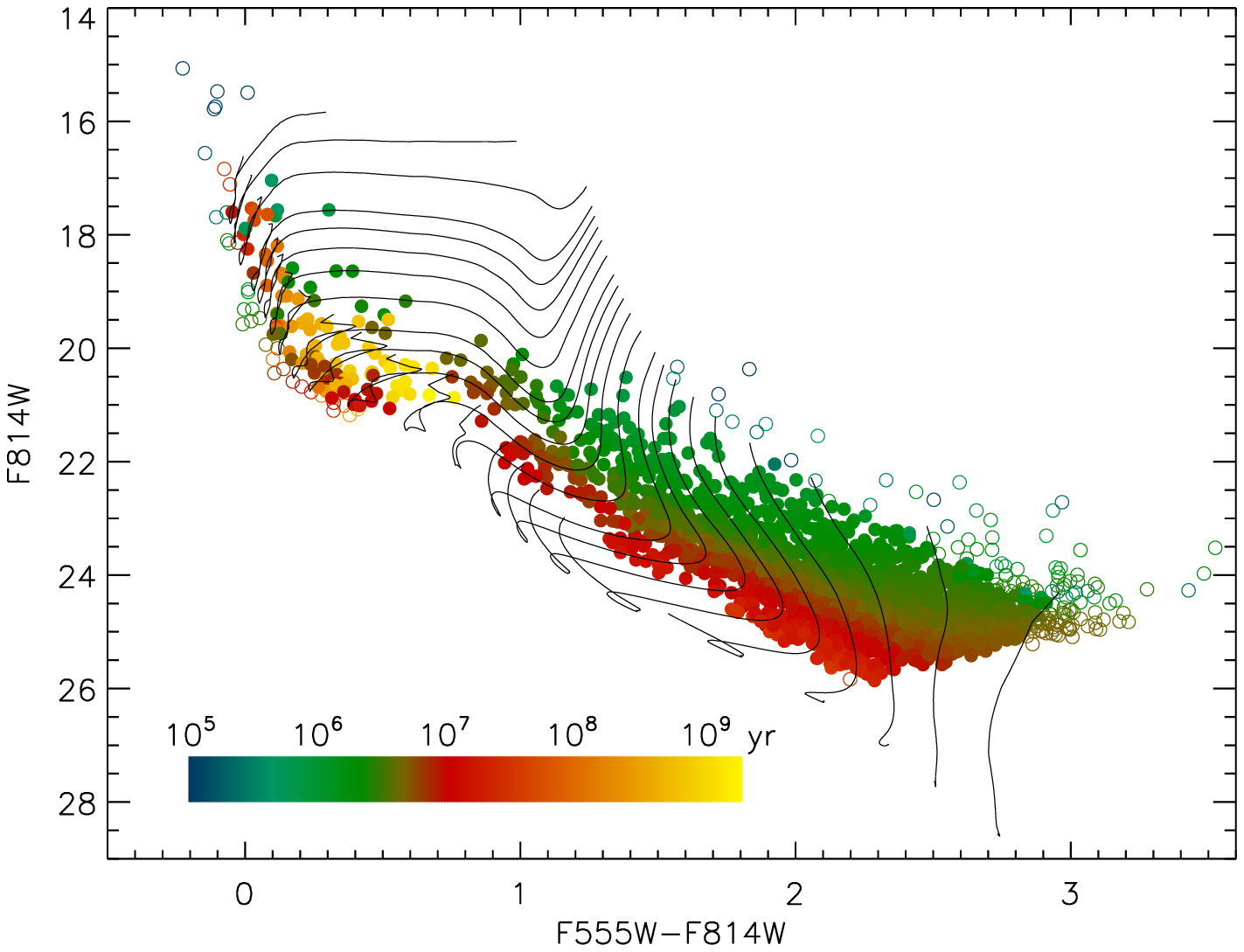}
\caption{PMS evolutionary tracks from \citet{tognelli2011} converted into the ACS photometric system using TA-DA, together with the observed photometry of LH~95. Sources are color coded according to their masses (\emph{left panel}) and ages (\emph{right panel}), as derived by the TA-DA fitter. Open squares are sources that the software identifies to be outside the range covered by the models. \label{figure:CMD-colorcoding}}
\end{figure*}
\begin{figure}
\epsscale{1.1}
\plotone{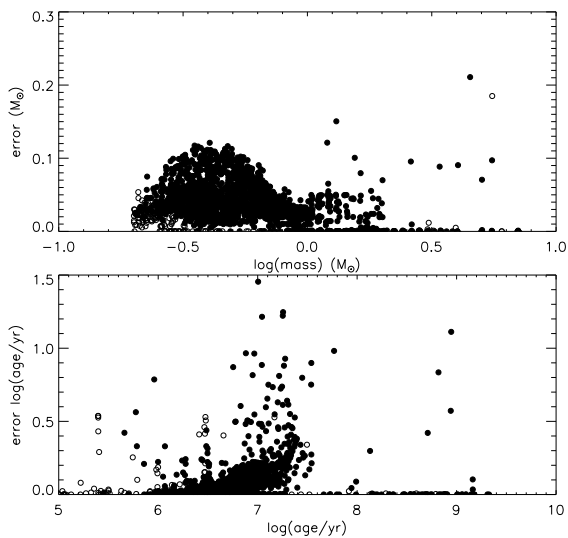}
\caption{Uncertainties in the LH~95 members masses and aged derived by TA-DA, as a function of these two quantities, originated from the individual photometric errors. \label{figure:CMD-errors}}
\end{figure}

\subsection{Non-photospheric fluxes}
\label{section:caveats-excesses}
As a probably unnecessary note, we highlight that TA-DA is meant for estimating stellar parameters in the case that the observed fluxes depend only the photospheric emission and presence of dust extinction. When this is not true, such as when the fluxes are contaminated by circumstellar emission or other flux excesses, the results obtained could be wrong. In some cases, however, it is possible to use TA-DA for non standard applications, such as disentangling photospheric properties from flux excesses. As an example, having at disposal infrared multi-band photometry, one can isolate the excess due to circumstellar disks around young stars and derive $T_{\rm eff}$ for individual stars. To do so, one could create a tuned grid of synthetic spectra in which instead of $\log g$ as second parameter (besides $T_{\rm eff}$), the amount of disk excess is added. Then TA-DA can be run on a multi-color observational space, and the fitter will provide the solution for both internal and external effects on the stellar flux.

\subsection{Degeneracies in the solutions}
\label{section:caveats-degeneracies}

In some cases one could encounter degeneracies between the stellar parameters. For example, when dereddening a CMD onto an isochrone, for some sources there could be multiple intersection between the de-reddening line and the model. Since the fitter of TA-DA simply minimizes the $\chi^2$, in case of multiple solutions, only one is selected. However, the Monte-Carlo simulation used to estimate the uncertainties will likely converge on different solutions in different iterations; this leads to a larger associated uncertainty for the parameters, and this can be used as a diagnostic tool when analyzing the results. If multiple solutions are caused by different intersections of the de-reddening line with the models, constraining appropriately the $A_V$ range allowed during the fitting can also solve the problem.

\begin{figure*}
\epsscale{1.1}
\plottwo{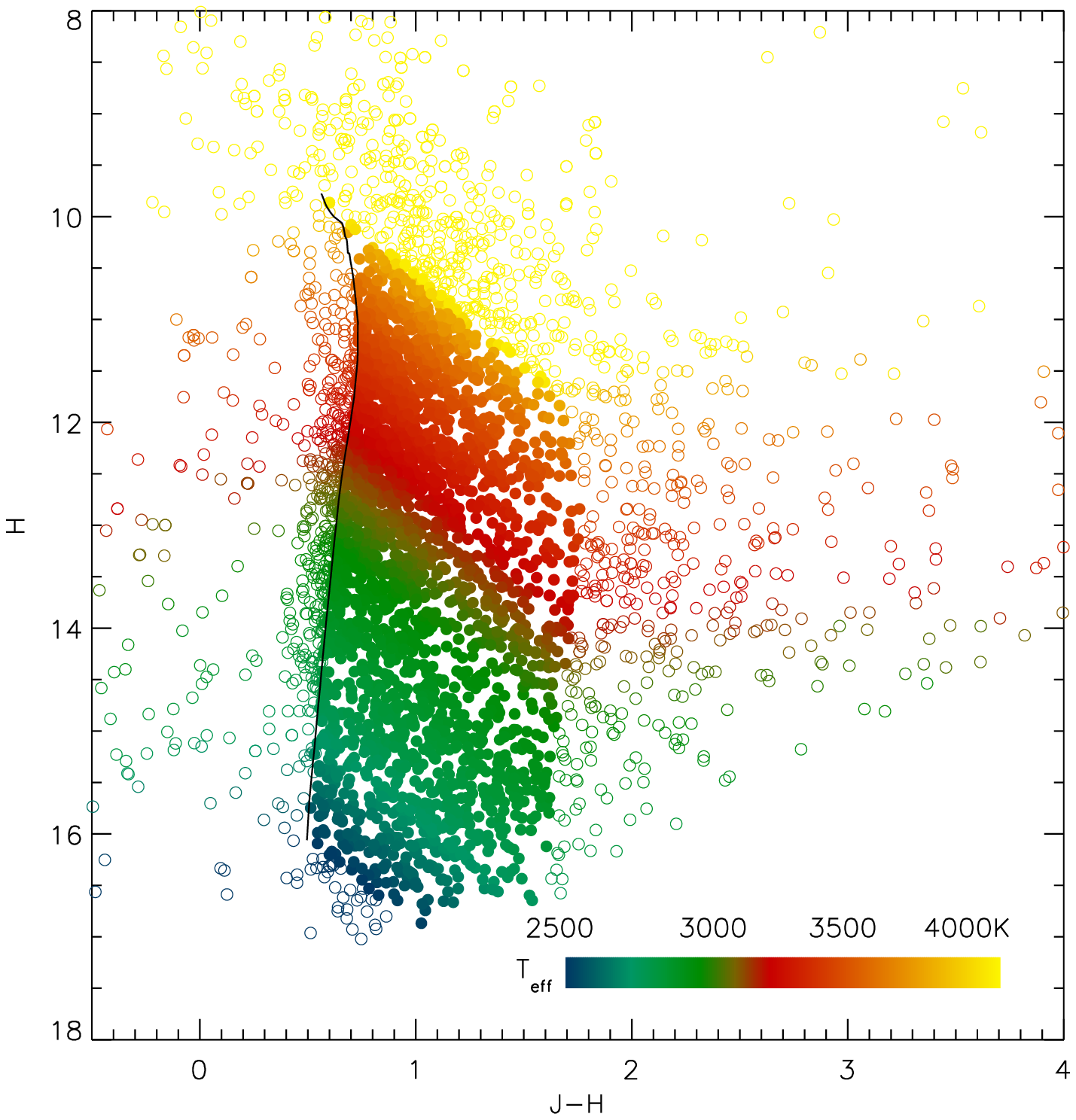}{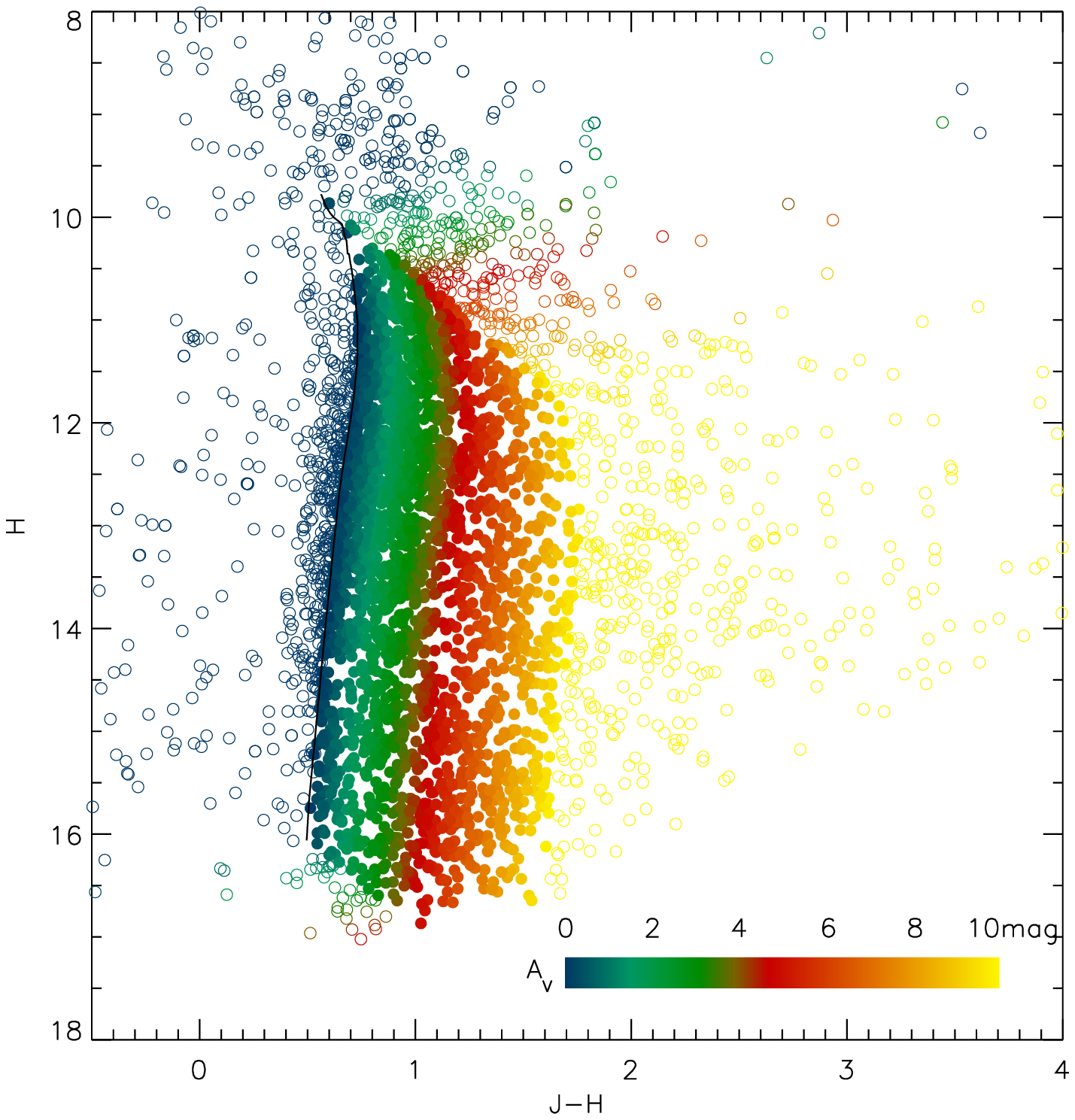}
\caption{Same as Figure \ref{fig:NIR}, for the second example of Section \ref{section:examples}. Here we consider NIR photometry of the ONC, and leave as free parameters the photospheric locus on a 2~Myr isochrone, as well as the extinction $A_V$, constrained to be from 0 to 10~mag. The data in the left and right panels are color-coded, respectively, for $T_{\rm eff}$ and $A_V$ as derived by the TA-DA fitting routine. \label{fig:NIR}}
\end{figure*}

\section{Examples of applications of TA-DA}
\label{section:examples}
We present here some practical examples of applications of TA-DA, for illustrative purposes.

\subsection{CMD interpolation}

Let's suppose we have a CMD and we aim at interpolating isochrones and tracks to estimate ages and masses of the sources. We consider the photometry from \citet{dario2009a}, obtained with the \emph{Hubble Space Telescope }(HST) \emph{Advanced Camera for Surveys} (ACS) on the young cluster LH~95 in the LMC, using the filters F555W and F814W. We run TA-DA, assuming the models from \citet{tognelli2011} with mixing length parameter $\alpha=1.68$, normal initial He and Deuterium abundance, and the LMC metallicity $Z=0.008$. We perform the synthetic photometry in our 2 filters with TA-DA considering the synthetic spectra BT-Settl from \citet{allard2011}, for the same metallicity as for the interior models; we assume a true distance modulus $\mu=(m-M)_0=18.41$, and two values of $A_V$: 0 and 0.5 (the average extinction for LH~95), considering the LMC average reddening law from \citet{gordon2003}. We save the synthetic photometry to file for later use, and we load the table with the observed photometry to TA-DA. We run the fitter, constraining $A_V=0.5$ and performing the MC simulation for the uncertainties. Results are shown in Figure \ref{figure:CMD-colorcoding}: here we show the models converted in our CMD plane, together with the sources color-coded according to their estimated parameters from TA-DA. It is evident how TA-DA correctly interpolates the models to the observations, deriving masses and ages. It also identifies the sources located outside the model grid, and assigns them the parameters of the closest model point. There is some scatter in the assigned ages for intermediate mass stars: this is due to a residual degeneracy in the models in this parameter range, where a given point in the CMD can be associated both to a PMS and post-MS evolutionary stage. In Figure \ref{figure:CMD-errors} we plot the errors in age and mass as a function of the derived parameters, due to the photometric errors, and computed by TA-DA using a MC simulation. The age uncertainty increases with ages, for the reason mentioned above. The increase in the mass uncertainty do the low-mass range, on the other hand, is due to larger photometric errors at faint luminosities.

\subsection{De-reddening to an isochrone}

In this second example, we consider a CMD and use TA-DA to deredden all the sources to a single given isochrone. We use the $JHK$ photometry of the Orion Nebula Cluster (ONC) presented in \citet{robberto2010}. We run TA-DA considering a single 2~Myr isochrone from the models of \citet{baraffe98}, we apply the distance modulus for the ONC of 8.085~mag, consider the BT-Settl models for solar metallicity, and obtain the synthetic photometry in the 2MASS photometric system (which is the one in which our data are expressed. We run the fitter leaving $A_V$ as a free parameter, in the range $0\leq A_V \leq 10$~mag. The result is illustratively shown in Figure \ref{fig:NIR}. As for Figure \ref{figure:CMD-colorcoding}, open dots represent sources outside the considered range of parameters; it can be noted that all the sources at red colors ($J-H\gtrsim2$) appear as such. This is because of the upper limit imposed for $A_V$. We also highlight that for these sources, the parameters (indicated by the color-coding of the data points) represent the ``closest'' model points, along the direction defined by the individual photometric errors in the $J,H$ 2-magnitude plane. This is the reason why, in the left panel of Figure \ref{fig:NIR} and for $(J-H>2)$, the estimated $T_{\rm eff}$ appears to decrease with increasing color $(J-H)$

\subsection{A multi-band example}

\begin{figure}
\epsscale{1.1}
\plotone{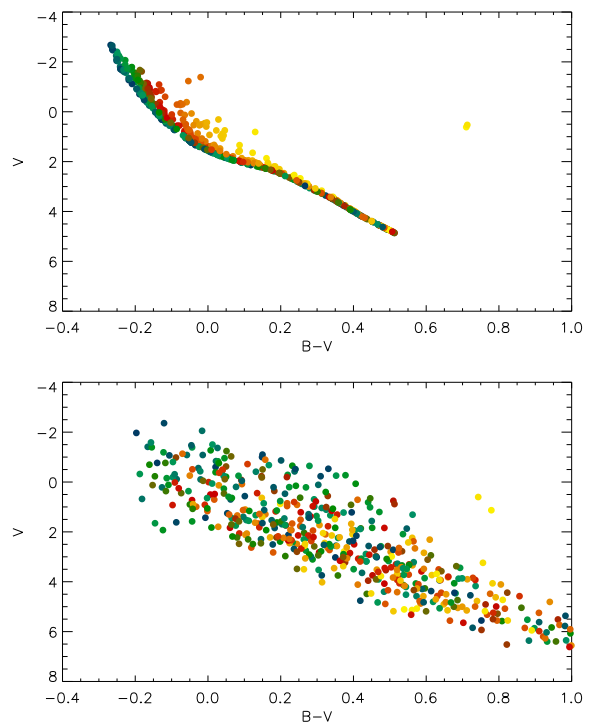}
\caption{Our simulated photometry of 500 stars with masses between 1 and 10~M$_\odot$ and ages between 10~Myr and 1~Gyr, in the $BV$ CMD. The upper panel shows the intrinsic magnitudes for $A_V=0$, color-coded according to the stellar age. The lower panel shows the same population after applying a random differential extinction in the range $0\leq A_V\leq 2$ and photometric errors. \label{fig:multifit-CMD}}
\end{figure}
\begin{figure*}
\epsscale{0.9}
\plotone{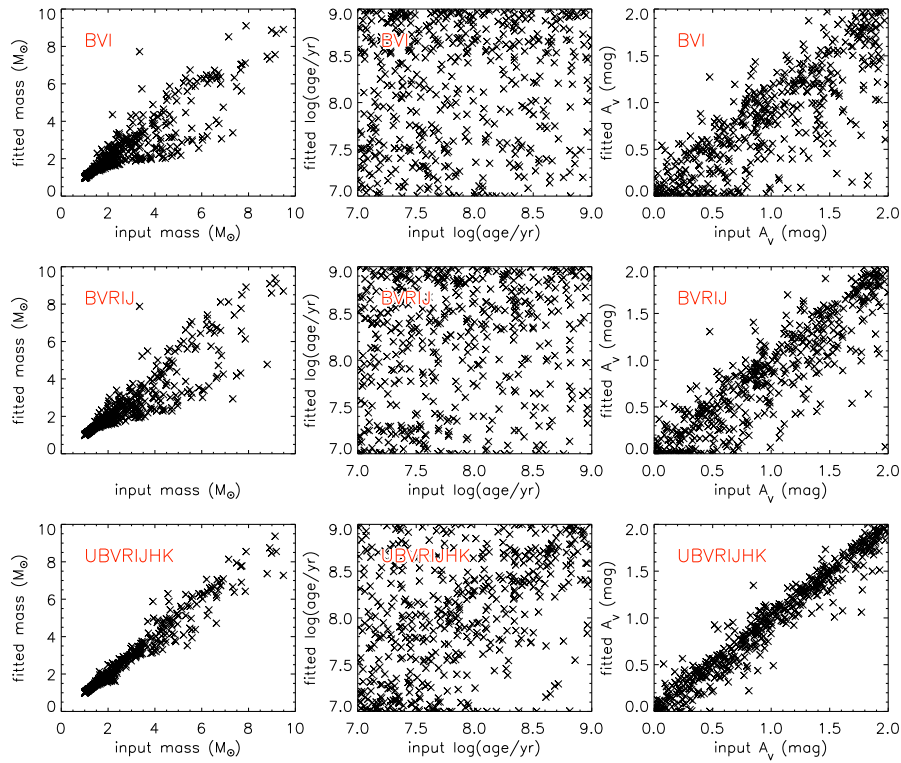}
\caption{Comparison of the original parameters of the 500 simulated stars, and the parameters recovered by the fitter within TA-DA after adding a random amount of extinction and photometric errors. The three rows shows the result of the fit using only 3, 5 or all 8 $UBVRIJHK$ bands. \label{fig:multifit-mags}}
\end{figure*}
\begin{figure}
\epsscale{1.1}
\plotone{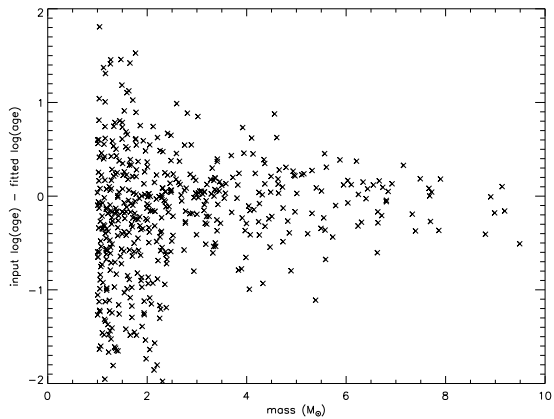}
\caption{The error in the estimated ages as a function of stellar mass. \label{fig:multifit-agetest}}
\end{figure}

In the next illustrative example we show the capabilities of TA-DA in dealing with multi-band photometry. We start by simulating a population of 500 evolved sources, with random masses in the range $1<M/{\rm M}_\odot<10$ and random ages between 10~Myr and 1~Gyr. We utilize TA-DA, assuming the models of \citet{marigo2008} first to compute the photospheric parameters ($T_{\rm eff}$, radius, etc.), then to derive by means of synthetic photometry their magnitudes in the standard $UBVRIJHK$ filters (Johnson, Cousins, and 2mass systems). An example of the result, limited to the $BV$ CMD, is shown in Figure \ref{fig:multifit-CMD}, upper panel. Next, we add a random amount of extinction to each of the 500 simulated stars, in the range $0<A_V<2$. We further displace the magnitudes by adding random, uncorrelated photometric errors to all the magnitudes, assuming an amplitude $\sigma=0.05$~mag. The result, again in the $BV$ CMD is show in Figure \ref{fig:multifit-CMD}, bottom panel; it is evident how at this point, a single CMD is insufficient to recover the properties of the individual sources, as they appear well mixed.

We now use the fitter integrated in TA-DA to try to recover the original parameters of these 500 stars. We consider the entire grid of interior models from \citet{marigo2008}, convert them in absolute magnitudes the $UBVRIJHK$ bands, load the simulated table with the photometry of our simulated sample, and run the fitter to estimate the most likely parameters of the stars, leaving all mass, age, and $A_V$ as free parameters. We perform this in multiple configurations, varying the number of photometric bands used to estimate the parameters. Some examples are shown in Figure \ref{fig:multifit-mags}. Here we compare the original mass, age and extinction, with those automatically derived by the fitter, using only 3 bands ($BVI$), 5 bands ($BVRIJ$) or all 8 bands ($UBVRIJHK$). It is evident that increasing the number of bands leads to a better estimate of the parameters; with all 8 bands, TA-DA is able to fully disentangle the differential $A_V$ from the stellar mass, thus recovering the properties of the sources. The residual scatter is due to the random photometric errors that we introduced.

The central column of Figure \ref{fig:multifit-mags} shows that, on the other hand, the estimate of the age remains highly uncertain. This is not a problem with the fitting procedure, since, as the extinction is correctly estimated, the original position of the sources in the HRD is fully recovered, rather, this indicates that ages are somewhat degenerate in the sampled parameter range. In fact, as evident from Figure \ref{fig:multifit-CMD}, upper panel, most of the stars are below the turn-off, and for the lowest masses, sources are located on the same main sequence regardless their age. Figure \ref{fig:multifit-agetest} confirms this: the error in the estimated ages (the difference between the best-fit age and the original age) is largest for the lowest masses in the sample, and progressively decreases towards the high-mass end.

\section{How to obtain TA-DA}
TA-DA can be retrieved from the following URL: \texttt{http://www.rssd.esa.int/staff/ndario/TADA/}. From this page the installation package, the manual, as well as additional grids of models can be downloaded.

\acknowledgements
TA-DA is developed thanks to funding from the \emph{National Aeronautics and Space Administration} (NASA) through award grant number NNX07AT37G. The authors thank Maddalena Reggiani,  Abhijit Rajan, Marc Postman, and Jes\'us Ma\'iz Apell\'aniz for their help in the development of this software, and the team of scientists that proposed the project.

\appendix
\section{Evolutionary models, synthetic spectra and filter throughputs available}
\label{appendix:models-filters}
TA-DA includes the following evolutionary models: for PMS stars the \citet{palla99}, the \citet{dantona98}, the \citet{baraffe98}, the \citet{siess2000} and the \citet{tognelli2011}. The first three of these are available only for solar metallicity, whereas the last two include different values of [M/H]. The \citet{tognelli2011} models, furthermore, include isochrones and tracks for several assumed values of the mixing length parameter $\alpha$, as well as different initial He and Deuterium abundances. For evolved population we integrated the \citet{marigo2008} isochrones.

\begin{deluxetable}{rl}[!h]
\tabletypesize{\tiny}
\tablecaption{Photometric filters included in TA-DA\label{table:filters}}
\tablehead{\colhead{Photometric system} & \colhead{Filter} }
\startdata
2MASS         & $J$, $H$, $Ks$ \\
BESSELL       & $U$, $B$, BW, $V$, $R$, $I$ \\
CFHT          & $I$, $Y$, $J$, $H$, $Ks$ \\
COUSINS       & $R$, $I$ \\
GALEX         & FUV, NUV \\
HST/ACS HRC   & F220W, F250W, F330W, F344N, F435W, F475W, F502N, F550M, F555W, F606W, \\
              & F625W, F658N, F660N, F775W, F814W, F850LP, F892N \\
HST/ACS SBC   & F115LP, F122M, F125LP, F140LP, F150LP, F165LP \\
HST/ACS WFC   & F435W, F475W, F502N, F550M, F555W, F606W, F625W, F658N, F660N, F775W, \\
              & F814W, F850LP \\
HST/NICMOS    & F110W, F160W, F165M, F187W, F190N, F205W, F207M, F222M \\
HST/STIS CCD  & 50CC, F28X50LP \\
HST/STIS FUV  & 25M, F25LY, F25Q, F25SRF2 \\
HST/STIS NUV  & 25M, F25CIII, F25CN182, F25CN270, F25MGII, F25Q, F25SRF2 \\
HST/WFC3 IR   & F098M, F105W, F110W, F125W, F126N, F127M, F128N, F130N, F132N, F139M, \\
              & F140W, F153M, F160W, F164N, F167N \\
HST/WFC3 UVIS & F200LP, F218W, F225W, F275W, F280N, F300X, F336W, F343N, F350LP, F373N, \\
              & F390M, F390W, F395N, F410M, F438W, F467M, F469N, F475W, F475X, F487N, \\
              & F502N, F547M, F555W, F600LP, F606W, F621M, F625W, F631N, F645N, F656N, \\
              & F657N, F658N, F665N, F673N, F680N, F689M, F763M, F775W, F814W, F845M, \\
              & F850LP, F953N, FQ232N, FQ243N, FQ378N, FQ387N, FQ422M, FQ436N, FQ437N, \\
              & FQ492N, FQ508N, FQ575N, FQ619N, FQ634N, FQ672N, FQ674N, FQ727N, FQ750N, \\
              & FQ889N, FQ906N, FQ924N, FQ937N \\
HST/WFPC2     & F170W, F255W, F300W, F336W, F336W-no-leak, F336W-leak, F380W, F439W, \\
              & F450W, F467M,F502N, F547M, F555W, F569W, F606W, F631N, F656N, F673N,  \\
              & F675W, F702W, F785LP, F791W, F814W, F850LP \\
ISAAC         & F1215, F1710, F2090, F3280, FLBB \\
JOHNSON       & $U$, $B$, $V$ \\
JWST/NIRCAM   & F070W, F090W, F115W, F140M, F150W, F150W2, F162M, F164N, F182M, F187N, \\
              & F200W, F210M, F212N, F225N, F250M, F277W, F300M, F322W2, F323N, F335M, \\
              & F356W, F360M, F405N, F410M, F418N, F430M, F444W, F460M, F466N, F470N, \\
              & F480M \\
LANDOLT       & $U$, $B2$, $B3$, $V$ \\
SDSS          & $u$, $g$, $r$, $i$, $z$ \\
SPITZER/IRAC  & CH1, CH2, CH3, CH4 \\
STROMGREN     & $u$, $u_{old}$, $b$, $v$, $y$ \\
TYCHO         & $B$, $V$ \\
WFI           & MB571, MB620, MB753, MB770, MB851, MB852, MB853, MB870, $U$, $U$841, $B$, $B$842, \\
              & $V$, $R$, $I$, $I$879, H$\alpha$ \\
WISE          & W1, W2, W3, W4 \\
...           & White throughput \\
\enddata
\end{deluxetable}

For what concerns the atmosphere models, TA-DA integrates the following grids: the \citet{kurucz93} models, the phoenix/NextGen models from \citet{hauschildt99}, the AMES-MT models from \citet{allard2000} (complemented with the NextGen grid for $T_{\rm eff}>5000$~K), the AMES-Settl from \citet{allard2001}, the BT-Settl models from \citet{allard2011}. These grids cover different ranges of $T_{\rm eff}$, surface gravity $\log g$, and metallicity. We refer the reader to the manual of TA-DA for these specifications.

TA-DA integrates a large number of band profiles for most of the commonly used photometry systems. These are listed in Table \ref{table:filters}; we highlight that, although TA-DA is natively oriented to the optical and NIR wavelength range (since it addresses mostly problems linked to stellar photospheric fluxes), filters spanning from the UV to the mid-IR have been included. In the latter case (see, e.g, the Spitzer/IRAC and the WISE filters), the fact that TA-DA natively allows one to consider flux units (Jansky or erg s$^{-1}$cm$^{-2}$$\AA^{-1}$) -- commonly used at long wavelengths -- besides magnitudes, simplifies the use of this tool.

We stress that additional interior models, synthetic spectra, and filter throughputs can be added to the tool by the user with no need to modify the code. This possibility opens to future expansions of the capabilities of TA-DA.

\section{A note on the interpolation of PMS isochrones in the HRD}
\label{appendix:isoch-interp}

\begin{figure}
\plotone{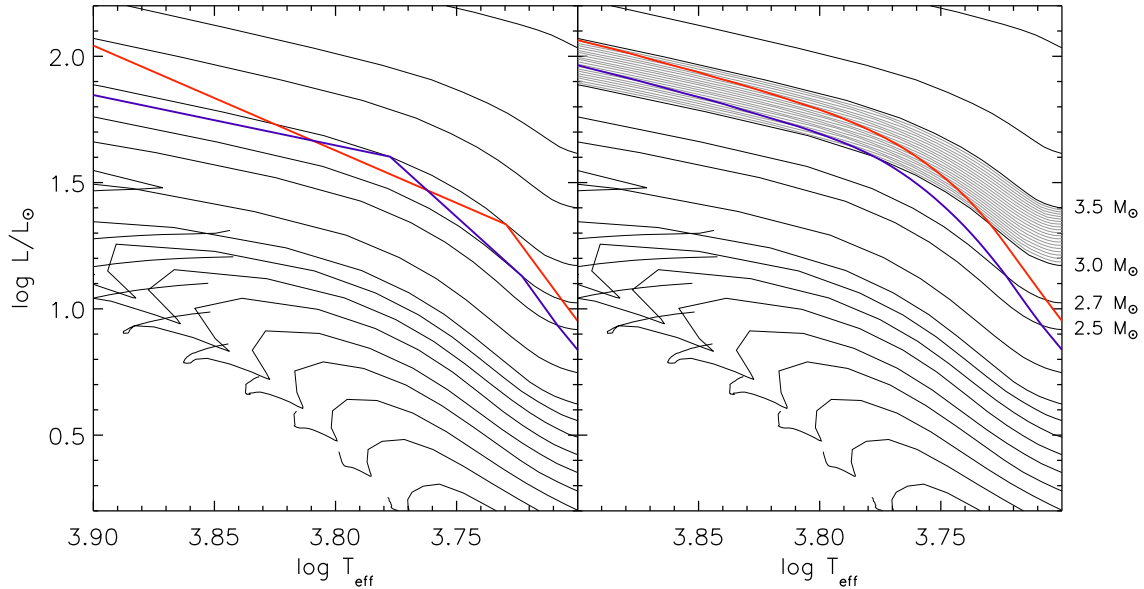}
\caption{A portion of the evolutionary tracks from \citet{siess2000}, for the PMS-MS turn-on at intermediate masses. The left panel shows two isochrones, (1.6 and 2~Myr), obtained by connecting the points with the considered ages on each original track. The right panel shows the same isochrones obtained from the improved interpolation technique described in the text. \label{fig:iso-interp}}
\end{figure}

It is not uncommon, especially for PMS models, to be computed with a sparse sampling of the stellar masses, and this can leads to a ``segmented'' appearance of the isochrones. This issue may become somewhat more serious in some ranges of the mass-age space, where the stellar evolution is very fast in the HRD and the rapidity of the evolution of $T_{\rm eff}$ and $L$ changes quickly for contiguous mass tracks. This is the case of the PMS-MS turn-on, as illustrated in Figure \ref{fig:iso-interp}, left panel. In this example we are showing the PMS tracks from \citet{siess2000}, as originally computed by these authors, and 2 isochrones for 1.6~Myr and 2~Myr, obtained in the standard way by connecting with straight segments the points, on all the tracks, corresponding to a given age. Between $\sim 2.5$ and $3.5$~M$_\odot$ the two isochrones cross, as it should not happen, and the segment connecting the 3.0 and 3.5~M$_\odot$ of the youngest isochrone passes in the region of the HRD between the 2.7 and 3.0~M$_\odot$ tracks.

Obviously, the ideal solution to this problem would be to have evolutionary models originally computed with a finer sampling of masses; when this is not possible one can only interpolate.
The correct way to perform this is to intepolate, from neighboring mass tracks, points corresponding to identical evolutionary stages. This is the common practice, e.g., to derive post-MS isochrones, where the timescales of different evolutionary stages are very different between them and highly dependent on the mass. For PMS stars, this method of interpolation is used, e.g., in the Pisa models from \citet{tognelli2011}.

Unfortunately, this is not always possible. For example the \citet{siess2000} PMS tracks are not provided for identical evolutionary points on different tracks. To improve on this, and compensate the issue shown in Figure \ref{fig:iso-interp}, left panel, we propose an alternative method to perform this interpolation, maximizing the information present in the models.
Our method is based on noting that, besides the 4 parameters listed in the evolutionary tracks (mass, age, $\log T_{\rm eff}$, $\log L$, contiguous mass tracks tend to have very similar shapes and sizes in the HRD, despite showing abrupt changes in the rapidity of evolution in the HRD. This is guaranteed by the fact that stars of similar mass undergo the same physical stages during their evolution, despite the fact that these may occur at very different times. Thus, we proceed at follows. First we interpolate new (interposed) mass tracks based in the shapes and lengths of the available ones. This can be done by assuming a metric in the HRD (e.g., a length $l\propto \sqrt{(10\log T)^2+\log L^2}$), divide each pair of contiguous mass tracks into a fixed, large, number (n$>$1000) of segments of identical length, and linearly interpolate all the 4 parameters listed above between corresponding pairs of points on both tracks. Qualitatively, our approach may be thought as a ``morphing'' technique, which progressively ``deforms'' one track into the next one based on their shape, size and position in the HRD.

The result is shown in Figure \ref{fig:iso-interp}, right panel. We plot 20 intermediate mass tracks interpolated between the original 3 and 3.5~M$_\odot$ ones, computed as described. These now nicely follow the shape of the two original ones. We also show the same 2 isochrones as in the left panel, now obtained connecting points of same age on all the intermediate interpolated tracks. These isochrones now never cross, and also appear much smoother in the HRD.

Some of the PMS models included in TA-DA \citep[specifically, the][]{siess2000,baraffe98,dantona98,palla99} have been pre-interpolated using the approach we described, creating very dense grids of masses and ages, generally including $>100,000$ points per model.

\end{document}